# Cell wall proteins: a new insight through proteomics

Elisabeth Jamet, Hervé Canut, Georges Boudart and Rafael F. Pont-Lezica*

Surfaces Cellulaires et Signalisation chez les Végétaux, UMR 5546 CNRS-Université Paul Sabatier, 24, chemin de Borde Rouge, BP42617, 31326-Castanet-Tolosan, France.
*Corresponding author: R. F. Pont-Lezica (lezica@scsv.ups-tlse.fr)

## Abstract

Cell wall proteins (CWP) are essential constituents of plant cell walls involved in modifications of cell wall components, wall structure, signaling, and interactions with plasma membrane proteins at the cell surface. The application of proteomic approaches to the cell wall compartment raises important questions: Are there technical problems specific to cell wall proteomics? What kinds of proteins can be found in *Arabidopsis* walls? Are some of them unexpected? What sort of post-translational modifications have so far been characterized in CWP? The purpose of this review is to discuss the experimental results obtained so far using protomics, as well as some of the new questions challenging future research.

## Glossary

**AGP (arabinogalactan proteins):** cell wall highly glycosylated HRGP that contain repetitive motifs such as (Ser, Thr, Ala)-Hyp-(Ser, Thr, Ala)-Hyp or (Ser, Thr, Ala)-Hyp-Hyp.
**AG (arabinogalactan)-peptides:** AGP that have a predicted mature protein backbone of 10 to 13 amino acid residues.
**BBE:** berberine-bridge (S)-reticulin:oxygen oxidoreductases.
**CWP:** cell wall proteins.
**Extensins:** cell wall structural HRGP, with numerous Ser-$Hyp_n$ ($n \geq 3$) motifs separated by Tyr-, Lys-, His- and Val-rich regions.
**GAP (glycosylphosphatidylinositol anchored proteins)**: proteins that are lipid-anchored to the external phase of the plasma membrane.
**GH (glycoside hydrolases):** enzymes that hydrolyze the glycosidic bond between two carbohydrates, or between a carbohydrate and a non-carbohydrate moiety.
**HRGP:** hydroxyproline-rich glycoproteins, *i.e.* extensins and AGP.
**Lectins:** structurally diverse proteins that bind to specific carbohydrates.
**LRR (leucine-rich repeat):** short sequence motifs having diverse functions and cellular locations, usually involved in protein-protein interactions.
**PME (pectin methylesterases):** enzymes that catalyze the de-esterification of pectin into pectate and methanol.
**Proteome:** the complete profile of proteins expressed, at a given time and environmental conditions, in a given organ, tissue, or cell.
**PRP (proline-rich proteins):** structural CWP rich in proline residues.
**Transcriptome**: the complete collection of transcribed elements of the genome.

## Towards a more comprehensive view of cell wall proteins

Since the first observations of plant cells by Robert Hooke in 1665, and until the 1980's, the cell wall was considered a rigid, static structure. In the past twenty years it has become evident that the cell wall is a dynamic organization essential not only for cell division, enlargement and differentiation (as is the animal extracellular matrix) [1, 2], but also acting in response to biotic and abiotic stress [3, 4]. It is also the source of signals for cell recognition within the same or between different organisms [5-7]. Cell walls are natural composite structures, mostly made up of high molecular weight polysaccharides, proteins, and lignins, the latter found only in specific cell types. Since present knowledge of cell wall polysaccharides has been recently reviewed [8, 9], we will focus on *Arabidopsis thaliana* cell wall proteins (CWP) that can be involved in modifications of cell wall components, wall structure, signaling, and interactions with plasma membrane proteins at the cell surface.

Molecular biology techniques and the complete sequencing of the *Arabidopsis thaliana* genome greatly contributed to the description of many CWP gene families and their transcriptional regulation [10-23]. It was found that most CWP are encoded by multigene families. Systematic transcriptomic approaches were combined with genetic analyses [8], but these, do not address the occurrence of alternative splicing or the post-translational modifications of the proteins. In addition, proteins can move in and out of complexes, modifying their functionality. This level of complexity cannot be tackled using transcriptomics alone [24]. Proteomics attempt not only to give a larger vision of the proteins present in a particular organ at a given stage of development, but also deal with some of these issues. Several recent reviews on plant proteomics have described the available methods in this area [24-26], and the application of proteomics to the study of cell walls [27, 28].

In the past three years, several groups used proteomics to identify CWP in different *A. thaliana* organs. These studies only assess which proteins are present, whereas their relative abundance remains unknown. A more accurate vision of the cell wall proteome is emerging: new CWP families and post-translational modifications, in addition to N- and O-glycosylations, are being described. This raises the question of estimating the number of proteins present in the cell wall of *A. thaliana*. The annotation of the *A. thaliana* genome shows that about 17% of the genome, *i.e.* 5000 genes, encodes proteins with a predicted signal peptide that targets them to the secretory pathway. The Cell Wall Genomic Group at Purdue University (cellwall.genomics.purdue.edu/families/index.html) has listed the *A. thaliana* genes involved in cell wall assembly and modification: among them, around 500 genes encode extracellular proteins. However, this number is too low because only the gene families encoding CWP with known biochemical functions are considered. If the CWP recently identified in proteomic studies, as well as the multiple forms of proteins produced by alternative splicing and post-translational modifications, are taken into account, a reasonable estimate will yield between 1000 and 2000 different proteins. In this review of cell wall proteomics in *A. thaliana*, we will analyze the results, discuss the emerging picture, underline the specific contributions of proteomics, and point out to new perspectives in the area.

## The hard to grasp cell wall proteome

In addition to the difficulties usually encountered in proteome analysis, such as protein separation and detection of scarce proteins [29], CWP present specific complexities. They are embedded in an insoluble polysaccharide matrix and interact with other cell wall components (Box 1), making their extraction challenging. The available cell wall proteomes include labile and weakly bound proteins (Box 1). Weakly bound CWP are extracted from purified cell walls with salts or chelating agents. Since labile proteins can be lost during the preparation of cell walls, they must be extracted from tissues by non-destructive techniques such as vacuum infiltration [30], or recovered from liquid culture media from cell suspension cultures or seedlings [31, 32]. As of yet, there is no efficient procedure to release CWP strongly bound to the extracellular matrix. Structural proteins, for instance extensins or PRP, can be cross-linked *via* di-isodityrosine bonds [33, 34]. Until now, extensins have been eluted with salts prior to their insolubilization from cell suspension cultures [35]. Another difficulty is the separation of polypeptides by classical two-dimensional gel electrophoresis. Most CWP are basic glycoproteins (Figure 1 in Box 1) and are poorly resolved by this technique [36]. More than 60 % of CWP have a pI value between 8 and 12.9, with a mean of 8.5 (Box 1). This amount includes only a few structural proteins, well known for their basic pI. Finally, HRGP are heavily glycosylated, they are difficult to detect on gels, and resistant to proteases. Such proteins require the development of specific methods of isolation and deglycosylation such as those recently used for synthetic extensins [34] and AGP [37].

## Box 1. CWP and their interactions with cell wall components

Plant cell walls are complex structures composed of polysaccharides and proteins. Current models describe the arrangement of their components into two structurally independent and interacting networks, embedded in a pectin matrix [5, 69]. Cellulose microfibrils and hemicelluloses constitute the first network; a second one is formed by structural proteins. In this review, we consider as CWP all proteins secreted into the extracellular space as well as proteins located at the interface between the plasma membrane and the cell wall. CWP identified in proteomic studies are listed in a CWP database (supplementary material). Three types of CWP can be distinguished, according to their interactions with cell wall components. CWP can have little or no interactions with cell wall components and thus move freely in the extracellular space. Such proteins can be found in liquid culture media of cell suspensions or seedlings or can be extracted with low ionic strength buffers. We call this fraction "labile proteins", most of them have acidic pI ranging from 2 to 6 (Figure 1). Alternatively, CWP might be weakly bound to the matrix by Van der Waals interactions, hydrogen bonds, hydrophobic or ionic interactions. Such proteins may be extracted by salts and most of them have basic pI ranging from 8 to 11 so that they are positively charged at the acidic pH of cell walls (Figure 1). Even though most of the cell wall polysaccharides are neutral, negatively charged pectins contain polygalacturonic acid that provides negative charges for interactions with proteins with a high pI. Such interactions would be modulated by pH, degree of pectin esterification, $Ca^{2+}$ concentration, and by mobility and diffusion coefficients of these macromolecules [70]. Finally, CWP can be strongly bound to cell wall components so that they are still resistant to salt-extraction. As examples, extensins are cross-linked by covalent links [33, 71] and peroxidases that can have a high affinity for $Ca^{2+}$-pectate [72].

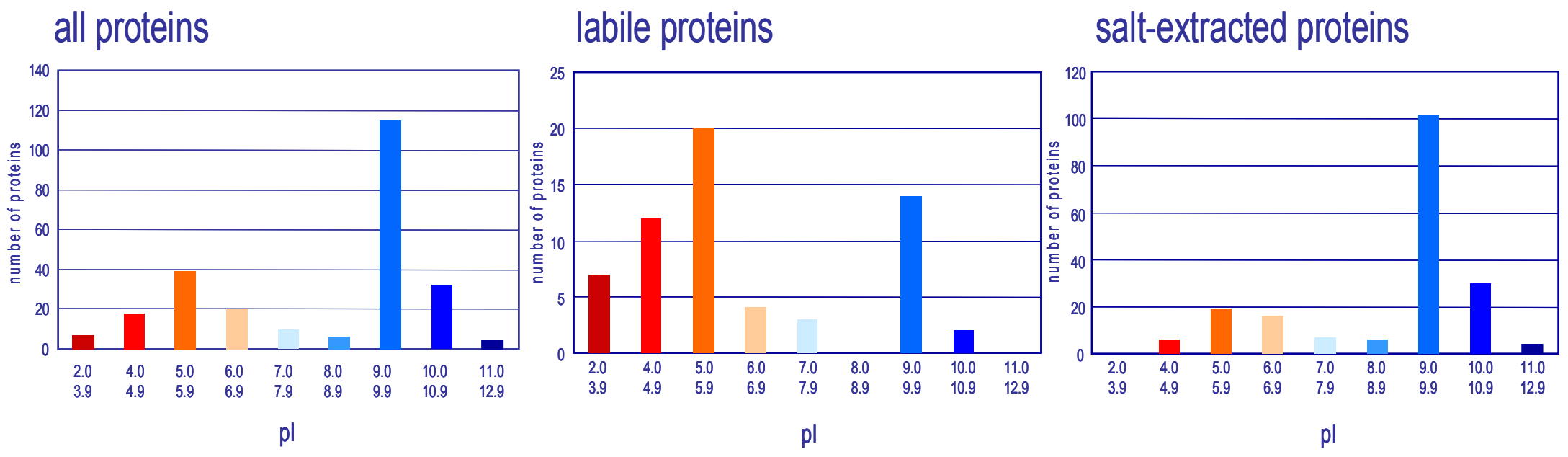

**Figure 1.** pI of CWP. pI of mature CWP were calculated after removal of their signal peptides (www.iut-arles.up.univ-mrs.fr/w3bb/d_abim/compo-p.html).

## Are there non-canonical CWP?

Non-canonical CWP, are known intracellular proteins. They have been reported in several recent publications on cell wall proteomes in higher plants [38, 39], green alga [40], and fungi [41, 42]. The presence of these proteins (*e.g.* enzymes of the glycolytic pathway, transcription factors, and ribosomal proteins) is puzzling, because they do not contain a predicted signal peptide, necessary for targeting to the secretory pathway; neither do they have an understandable function in the wall. The existence of an unknown export mechanism should not be excluded [43]. In these studies, non-canonical CWP can represent half of the proteins identified in cell wall preparations [38, 39]. They were extracted from isolated cell walls, and the results are highly dependent on the reliability of purification techniques. Indeed, these proteins, notably the basic ones, can be ionically trapped by the acidic polysaccharide matrix during cell wall purification. The purity of cell wall preparations can be checked using marker enzymes or antibodies against known proteins [38], but the analysis tool (mass spectrometry) is 10 to 1000 times more sensitive than classical biochemical and immunological tests. Using non-destructive techniques to isolate CWP [30], only a few non-canonical proteins were found, supporting the idea that they are likely to be contaminants.

## What's new on canonical CWP?

For this review, all available *A. thaliana* cell wall proteome data were screened using bioinformatic tools to select only those proteins containing a signal peptide but devoid of known retention signals for the endomembrane system [31]. These proteins were added to our CWP database which now has 281 proteins (supplementary material) [30-32, 38, 44-46]. In addition, the functional annotation was checked using both protein sequence comparisons and bioinformatic software designed to screen for functional domains [31, 47].

About 90 % of CWP were placed in categories on the basis of predicted biochemical or biological functions (Table 1). It should be noted that the biochemical function of only a small portion of the identified proteins was experimentally demonstrated. The assumption is that proteins sharing conserved domains have the same activity. The biggest surprise was that only half of the proteins had already been characterized as CWP. They are glycoside hydrolases (GH), carbohydrate esterases/lyases, expansins, oxido-reductases, structural proteins and proteins involved in signaling, of which most are AGP. The other half was partially known, *e.g.* some proteases and lectins had been described as being extracellular [9, 48, 49]. Most intriguing is the remaining 10 % of proteins that do not have any similarity to known proteins in other organisms. The challenge is to elucidate their biological role within the cell wall.

The CWP database (supplementary material) constructed for this review includes each protein's source of identification. Are the same proteins found in the cell walls of different organs? To try to answer this question, we have selected three sets of data, related to: leaves of fully developed rosettes containing differentiated cells, etiolated hypocotyls analyzed at the end of elongation, and 7-day-old cell suspension cultures, when cells are actively dividing and expanding. All three proteomes were obtained using comparable salt-extraction protocols, separation of proteins by electrophoresis, and identification by MALDI-TOF spectrometry. The comparability is based on identifying a given protein in a particular organ. Differences in polysaccharide composition, cell wall structure, a lower abundance of the protein, or post-translational modifications might lead to its not being detected during the experiment.

**Table 1. Predicted functional classes of labile and weakly bound CWP.**

The 281 proteins from the data base (supplementary material) were assembled in functional classes and subclasses according to the presence of functional domains in the protein. Proteins acting on polysaccharides include glycoside hydrolases, esterases, lyases and expansins; oxido-reductases include peroxidases and berberine bridge enzymes; proteins with interacting domains include proteins with lectin or LRR domains and enzyme inhibitors. Data originate from rosettes, cell suspension cultures, etiolated hypocotyls and seedlings, culture media of cell suspension or etiolated seedlings, and protoplasts.

| Functional classes | | % of identified proteins |
|---|---|---|
| **Proteins acting on polysaccharides** | | **29,5 %** |
| Glycoside hydrolases | 21 % | |
| Esterases/Lyases | 5.5 % | |
| Expansins | 3 % | |
| **Oxido-reductases** | | **13.5 %** |
| Peroxidases | 6 % | |
| Berberine bridge enzymes | 2.5 % | |
| **Structural proteins** | | **1.5 %** |
| **Proteins involved in signaling** | | **8 %** |
| **Proteases** | | **10 %** |
| **Proteins with interacting domains** | | **10 %** |
| Lectin domains | 2.5 % | |
| LRR domains | 3.5 % | |
| Enzyme inhibitors | 3 % | |
| **Miscellaneous** | | **16.5 %** |
| **Unknown function** | | **10 %** |

The number of CWP identified in each proteome and the number of common CWP are shown in Figure 1a. An interesting fact is the presence of eleven proteins common to all three organs. A closer look at those proteins reveals: two GH, two PME, one germin, one protease, and four proteins with interacting domains (three lectins and one homolog to a tomato xyloglucan-specific endoglucanase inhibitor protein), and one protein of unknown function. Interestingly, one of the two GH is an alpha-xylosidase encoded by a single gene (*At1g68560*). The question is: Do these CWP represent a group of housekeeping proteins, essential to all types of cell walls?

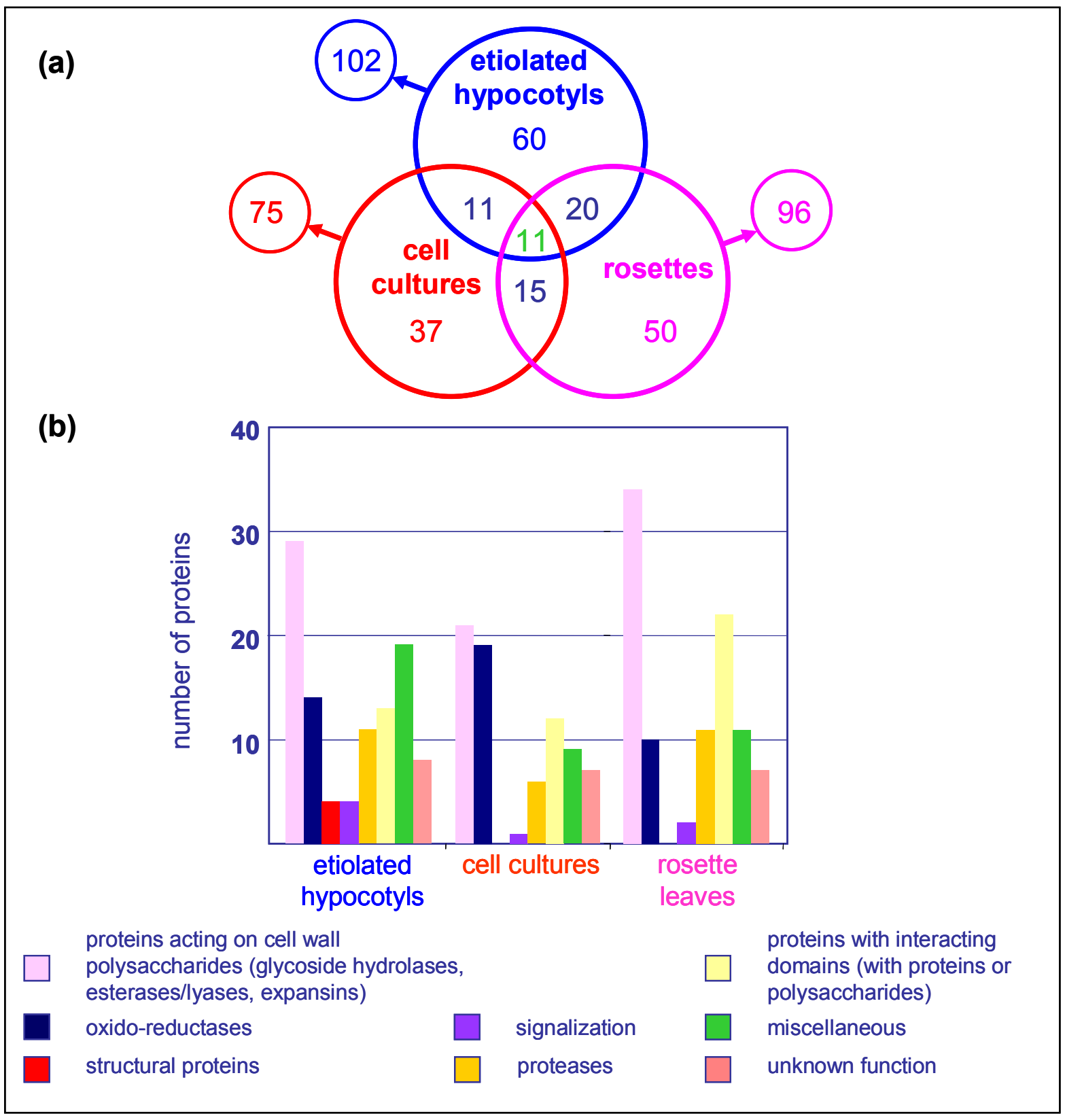

**Figure 1.** Comparison of partial cell wall proteomes of etiolated hypocotyls, cell suspension cultures, and rosette leaves. Labile and weakly bound CWP were represented. (a) Venn diagram illustrating the overlap between the three cell wall proteomes. (b) Histogram showing the functional classes of labile and weakly bound CWP in the different organs. Functional classes are as in Table 1 and all proteins are listed in the CWP database (supplementary material).

At least 50 % of the identified CWP of one proteome were not found in the others, and thus might be specific to that type of cell walls. This is partially linked to the high number of genes found for each CWP family, which might be differently regulated during development. A more detailed analysis (Figure 1b), showing the classes of CWP in each organ, confirms that in all cases the best represented proteins are those acting on cell wall polysaccharides. As expected from the fact that GH represents 20 % of the identified CWP (Table 1), proteins acting on cell wall polysaccharides also are the category with the highest diversity within each organ. Oxido-reductases are particularly numerous in cell suspension cultures, probably due to the mechanical stress produced by the continuous spinning, and to the oxidative stress that occurs in liquid media culture. The only organ in which a few salt-extractable structural proteins were identified is etiolated hypocotyls, possibly because such proteins are not yet completely insolubilized. Proteins having domains of interaction with proteins or polysaccharides are well-represented in all organs, and especially in rosettes.

## Well-known CWP: the lessons of proteomics

Proteomic analyses provide a great advantage: they allow for the precise identification of proteins belonging to the same family, and the detection of each member in different organs. As an example, Figure 2 shows the number of members of each GH subfamily found in rosette leaves, etiolated hypocotyls and cell suspension cultures. GH have been classified according to the CAZy nomenclature (afmb.cnrs-mrs.fr/CAZY/) based on sequence homology. In rosettes and etiolated hypocotyls, 23 and 17 GH have been identified respectively, whereas only 13 in cell cultures. The unexpected high number of different GH, in mature leaves suggests that their cell walls undergo constant change. In contrast, the low number found in cell cultures is surprising, because GH were expected to be well represented in dividing and elongating cells that are remodeling their polysaccharides. However, this low number can also be due to the absence of cell differentiation in cell cultures. Each organ seems to have a particular distribution of GH (Figure 2). It would be interesting to link these enzymatic activities to their substrates for a better understanding of polysaccharide remodeling in cell walls during cell division, cell expansion and differentiation. The same differences in GH distribution among different organs are found in other protein families, such as oxido-reductases.

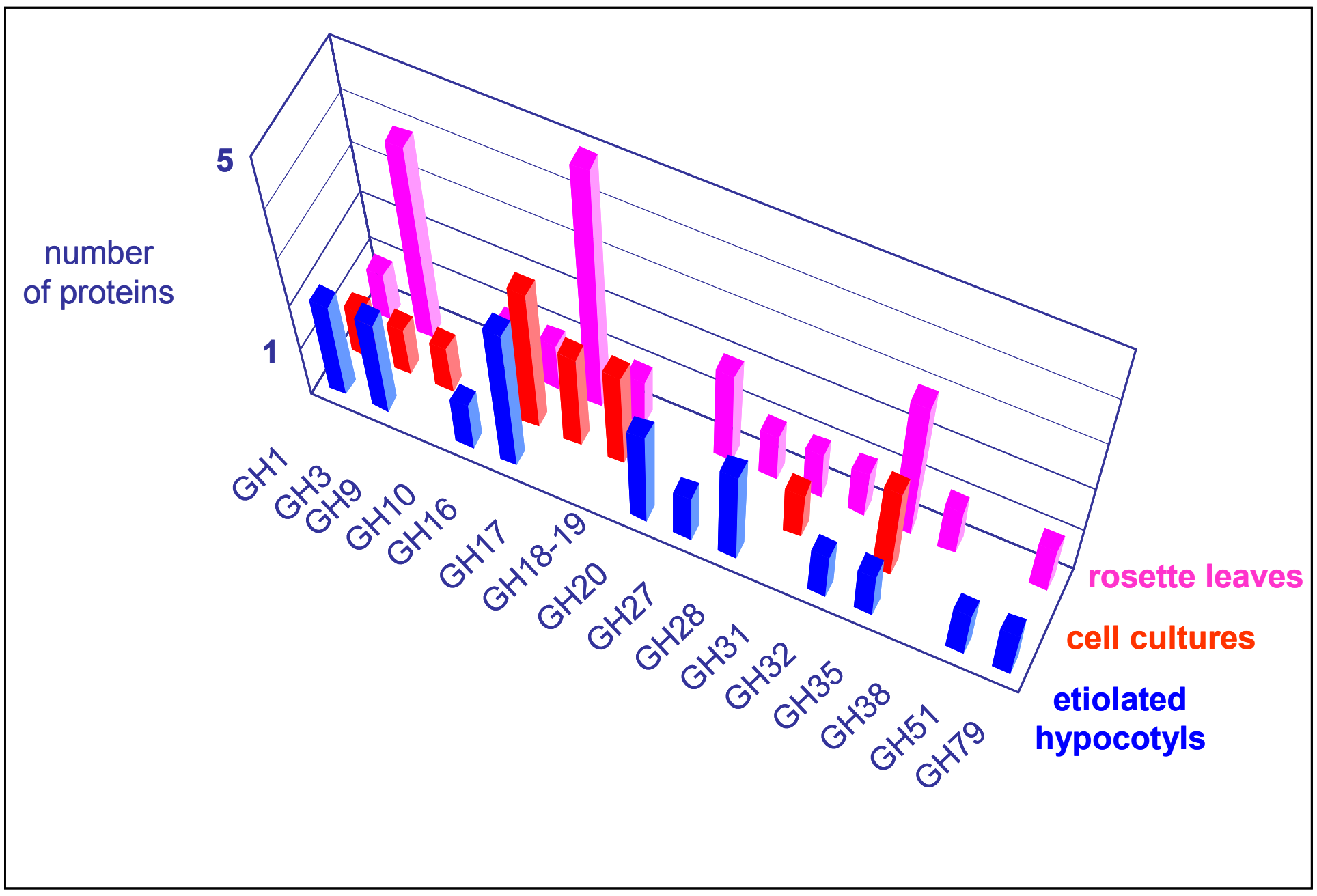

**Figure 3.** Occurrence of glycoside hydrolases in three cell wall proteomes. Data originate from etiolated hypocotyls, 7 day-old cell suspension cultures and rosette leaves, as listed in the CWP database (supplementary material). Glycoside hydrolases were classified according to the CAZy nomenclature (afmb.cnrs-mrs.fr/CAZY/), based on sequence homology.

Another feature of proteomics is the characterization of protein structure through the analysis of post-translational modifications. Glycosylation, hydroxylation, as well as many other modifications, are essential because they determine structure, localization, and activity. Such characterization of CWP is exemplified by glycosylphosphatidylinositol-anchored proteins (GAP), among which AGP [37, 45]. The properties of these proteins allowed specific purification procedures, and their separation as a subset of CWP. It was found that various protein families were lipid-anchored to the plasma membrane via a glycosylphosphatidylinositol (GPI) (supplementary material). The characterization of the peptide moiety of arabinogalactan (AG)-peptides belonging to a subfamily of AGP led to experimental evidence of the existence of GPI anchors, the determination of the cleavage site for both the endoplasmic reticulum secretion signal, and the GPI anchor signal for 8 of the 12 AG-peptides. Moreover, a new post-translational modification was found in AG-peptides, namely the hydroxylation of proline within the Gly-Pro motif [37].

## Towards new biological functions in cell walls

Proteome studies open new perspectives by expanding our knowledge of CWP. The identification of the different types of proteins contributing to the same physiological process should also help to better understand cell wall functions, as shown by the following examples.

Some proteases were already known to be localized in cell walls by immunological approaches and enzymatic activities [9, 50]. Proteomics provide additional information on the great diversity of proteases, such as homologs of subtilisin, carboxypeptidases, aspartyl, and cysteine proteases. So far, the *A. thaliana* mutant *sdd1-1* (stomatal density and distribution1-1) is the only described mutation affecting a gene encoding a cell wall protease [49]. The stomatal pattern is disrupted in the mutant, resulting in stomata clustering and increased stomatal density. However, overexpression of *SDD* produces the opposite phenotype. The authors propose that SDD1 generates an extracellular signal regulating the number of asymmetric divisions in satellite meristemoids. Proteases can consequently play a role in the generation of signals involved in development; they may also contribute to CWP turnover, a process still poorly understood.

Proteins with interacting domains such as lectins or LRR (leucine-rich repeats) proteins are likely to play essential roles in cell walls. Indeed, three lectins were found in all the cell wall proteomes discussed here, and carbohydrate recognition is important in self- and non-self interactions in plant cell walls. Interesting evidence is provided by the study of the zygote secreted (ZSP)-2 protein of *Chlamydomonas reinhardtii,* which has both lectin and HRGP domains [51]. The authors propose that ZSP-2 binds sugar residues to favor the assembly of the zygote cell wall. Thus lectins may be essential to the organization and assembly of the polysaccharide matrix. LRR-containing proteins are thought to interact with other proteins. In particular, a polygalacturonase-inhibiting protein (PGIP) has been well characterized: its structure has been established by X-ray crystallography [52] and its differential affinity for fungal polygalacturonases has been related to its LRR domains [53]. Furthermore, it was shown that several LRR-receptor kinases had important roles in development or disease resistance [53]; many are thought to participate in signaling. However, information on their ligands and ligand-binding sites is still largely lacking.

Redox reactions play many roles in plant cell walls during development as well as in response to pathogen attacks [54-57]. In addition to peroxidases, several CWP might be involved in such processes. For example, homologs of berberine-bridge (S)-reticulin:oxygen oxidoreductases (BBE) from *Papaver* and *Berberis* were found in the cell wall. Classical BBE are localized in vacuoles, and involved in the synthesis of alkaloids [58]. However, the *A. thaliana* proteins were predicted to be extracellular [31], and recently, a secreted tobacco BBE was found to have glucose oxidase activity [59]. The substrate specificities of cell wall BBE are therefore different from those of vacuolar proteins. Germins and germin-like proteins constitute a large and diverse family of ubiquitous plant proteins. In cereals, they were described as oxalate oxidases, strongly associated to hemicelluloses, the synthesis of which is linked to the increase in cell wall extensibility [60]. In several dicot germin-like proteins, this activity could not be assessed. However, a cotton germin-like protein was found to accumulate in the fiber apoplast during cell elongation [61, 62]. Finally, phytocyanins, classified as 'miscellaneous' in our database, and also known as blue copper proteins, may be associated, along with small molecular weight compounds, as electron transfer proteins in redox processes [59]. In addition to a copper binding domain, stellacyanins and uclacyanins have a cell wall structural protein domain, which suggests possibilities of associations with other structural proteins.

## The challenge of the newcomers in cell walls

Proteins with unknown functions are one of the greatest challenges in CWP groups. From our experiments, it seems that some of these proteins are major components of the cell wall proteome. It should also be noted that one of the unknown function CWP was common to the three analyzed proteomes. Specialists in protein structure have defined some "domains of unknown function" (DUF) shared by several protein families (www.sanger.ac.uk/Software/Pfam/search.shtml). DUF do not display homology with any domain of known function, and some of them are specific to plant proteins. The fact that some of these proteins can only be found in plants, that they could be abundant in cell walls, and that they have no known function, make them a target of choice for future studies.

## Concluding remarks

In this review, we have shown the multiple contributions of proteomics to the knowledge of CWP. Proteome analyses allow for: i) a precise identification of members of CWP families in specific organs; ii) the identification of new CWP (among which proteins of unknown function); iii) the characterization of CWP by studying post-translational modifications; iv) an overview of all the proteins present in cell walls at a particular physiological stage. This is, however, the tip of the iceberg, and a big effort should be made to increase the number of identified CWP. Additional information on CWP can be obtained using methods already developed in proteomic approaches for improving the extraction of CWP strongly bound to cell wall components, such as polysaccharides or lignins, and/or the separation of CWP prior, to identification by mass spectrometry [63, 64]. Since quantitative data are still missing, accurate comparisons between samples can be pursued by applying available techniques to perform differential proteomics [65, 66]. The understanding of the biochemical and/or biological functions of proteins, increasingly calls for the fine characterization of CWP [37, 67]. CWP interact with other cell wall components, including CWP or plasma membrane proteins. Such interactions could be studied using the BIA (biomolecular interaction analysis)-MS technology [68]. All these data will provide a better knowledge of CWP that combined with genetics, biochemistry, and molecular biology, can lead to understand the roles of CWP in plant development, signaling, defense, and adaptation to the environment.

## Acknowledgments

The authors are grateful to the Université Paul Sabatier (Toulouse, France) and the CNRS for support. They are indebt to Lorena Pont-Lezica and Diana Mosovich for their kind revision of the manuscript.